\documentclass[a4paper, final, 12pt]{article}
\usepackage{eurosym}
\usepackage{amsmath, amssymb, latexsym, amscd, amsthm,amsfonts,amstext}
\usepackage[mathscr]{eucal}
\usepackage{graphicx}
\usepackage{subfig}
\usepackage{float}
\usepackage{color}
\usepackage{hyperref}
\usepackage[utf8]{inputenc}
\usepackage[english]{babel}

\numberwithin{equation}{section}
\begin{document}

\title{An axially symmetric stationary N-center solution of Einstein's vacuum equations  }

\author{Aleksandr A. Shaideman$^{\ast }$, 
 Jesus D. Arias H. $^{\ast \ast  }$ and Kirill V. Golubnichiy$^{\circ }$ \\
%EndAName
 \and $^{\ast }$Department of Theoretical Physics
, \and  Peoples' Friendship University of Russia, \\ 
6 Miklukho-Maklaya St., 117198, Moscow,
 Russia  
 \and $^{\ast \ast }$ Dynamical Systems Research Group. 
\and Depatment of Mathematics and Physics.
\and Faculty of Basic Science and Engineering. University of Los Llanos.
\and Km 12 via Puerto Lopez, Vereda Barcelona.
 \and $^{\circ }$Department of Mathematics 
 \and Texas Tech University, Lubbock, TX 79409, USA  
 \and Emails: ashaideman@rambler.ru,
 \and jariar@unillanos.edu.co
\and kgolubni@ttu.edu }
\date{}
\maketitle

%\nocite{Strikwerda}
%\nocite{Tikhonov1943Stability}
%\nocite{KlibanovBuk}
%\nocite{Buk}
%\nocite{Klib1}
%\nocite{Bakush}
%\nocite{KlibLi}
%\nocite{BouIS}
%\nocite{LavR}

\begin{abstract}
Using the Euclidon method \cite{1}-\cite{3}, a stationary solution of Einstein's vacuum equations was obtained, describing N rotating axially symmetric masses, which in the absence of rotation describes N arbitrary axially symmetric static masses, for example, N Zipoy masses   on the axis of symmetry, and in the absence of distortion, N Kerr-NUT solutions.
\end{abstract}

%\newpage

\vspace{0.5em} %\textbf{\text{Notations}}

%\begin{itemize}
%   \item $\qtt \equiv (s_b, s_a) \times (0, 2\tau) \subset \mathbb{R}^2$.
%  \item $\partial \qtt \equiv \{(s, t): t = 0, (s = s_a \text{ or } s = s_b)\}$.
% \item $\mathcal{L}: H^{2, 1}(\qtt) \rightarrow L^2(\qtt)$ is the differential %operator, where
%    \[
%   \mathcal{L} u \equiv \frac{\partial u}{\partial t} + \frac{\sigma^2(t)}{2} s^2 %\frac{\partial^2 u}{\partial s^2}.
%   \]

%  \item $J_\beta(u) \equiv \int_{\qtt} (\mathcal{L}u)^2 dsdt + \beta \norm{u - %F}_{L^2(\qtt)}$ is the Tikhonov-like functional. 
%  \item $\bar{J}_\beta(u)$ is the discrete version of $J_\beta(u)$ 

%\end{itemize}

%\textbf{\text{Keywords:}}
%Vacuum stationary axially-symmetric Einstein equations, the Papapetrou-Weyl %metric, the Kerr-NUT solution,  stationary euclidon solution, asymptotically %flat solutions  

\section {Introduction}
     The first exact solutions of the axially symmetric stationary vacuum equations were presented by Lewis \cite{4} and Van Stockum \cite{5}. Later Papapetrou  \cite{6} obtained a canonical form of the line element, and Ernst \cite{7}  formulated the stationary vacuum problem in the form of a single equation for a complex function.                
     An outstanding result was obtained  in 1963 by Kerr \cite{8} , whose solution possibly describes the exterior gravitational field of a rotating source.
     A family of a asymptotically flat solutions of the stationary vacuum  equations was found by Tomimatsu and Sato \cite{9}, \cite{10}.
     Later, a survey of  known solutions was given by Kinnersley \cite{11} . Possible generalizations of the Tomimatsu-Sato metric were found by Yamazaki   \cite{12}, \cite{13} . Hori \cite{14} and Cosgrove \cite{15}, \cite{16}, and asymptotically not-flat solutions of the Ernst problem  were obtained in \cite{17}, \cite{19}, \cite{4}.
     The Ernst  equation \cite{7} makes it possible to extract  the solutions  by the method of separation of variables \cite{20}, \cite{21} . In a series of papers \cite{22}, \cite{23}, \cite{24}, a non-canonical form  of the stationary metrics  was used to obtain new solutions of the Einstein equations.
    An approximate method  for constructing solutions depending  on two harmonic functions was proposed  in \cite{25}, \cite{26}. Ernst \cite{27} dreamt of construction of such a kind of solutions, but exact ones. 
    
\section{Author declarations}
    The authors have no conflicts to disclose.
    
\section{Basic equations}
As was shown  by Papapetrou \cite{6},  the line element, describing a stationary axially symmetric gravitational field without loss of generality can be present in the following canonical form:
\begin{equation}
ds^{2} =f^{-1}[e^{2\gamma}(d\rho^{2}+dz^{2})+\rho^{2}d\varphi^{2}]-f (dt-\omega d\varphi)^{2}\label{3.1}
\end{equation}
Here $\rho, \varphi, z$ and $t$ are canonical Weyl coordinates and time, respectively; $f(\rho, z)$,  $\gamma(\rho, z)$ and  $\omega(\rho, z)$ are  unknown functions to be determined from the  field equations. 

The  vacuum Einstein equations are given by

\begin{equation*}
R_{ik}=0,
\end{equation*}

where $R_{ik}$ is the  Ricci tensor .

Therefore we obtain all  the Einstein equations in the case  of a stationary axially symmetric gravitational field  outside  the sources:

\begin{equation}
f\Delta f =(\overrightarrow{\nabla}f)^{2}-f^{4}\rho^{-2}(\overrightarrow{\nabla}\omega)^{2}, \label{3.2}
\end{equation}

 \begin{equation}
\overrightarrow{\nabla} (f^{2}\rho^{-2}\overrightarrow{\nabla}\omega)=0, \label{3.3}
\end{equation}

  \begin{equation}
\frac{\partial \gamma}{\partial \rho}=\frac{1}{4}\rho f^{-2}[(\frac{\partial f}{\partial \rho})^{2}-(\frac{\partial f}{\partial z})^{2}-f^{4}\rho^{-2}((\frac{\partial \omega}{\partial \rho})^{2}-(\frac{\partial \omega}{\partial z})^{2})], \label{3.4}
\end{equation}

  \begin{equation}
 \frac{\partial \gamma}{\partial z}=\frac{1}{2} \rho f^{-2}(\frac{\partial f}{\partial \rho}\frac{\partial f}{\partial z}-f^{4}\rho^{-2}\frac{\partial \omega}{\partial \rho}\frac{\partial \omega}{\partial z}), \label{3.5}
\end{equation}
 
The operators  $\overrightarrow{\nabla}$  and $\Delta$  defined by the formulae

\begin{equation*}
\overrightarrow{\nabla}\equiv \overrightarrow{\rho}_{0}\frac{\partial}{\partial \rho}+\overrightarrow{z}_{0}\frac{\partial}{\partial z}, 
\end{equation*}

\begin{equation*}
\Delta \equiv  \overrightarrow{\nabla}^{2} \equiv \frac{\partial^{2}}{\partial \rho ^{2}}+\frac{1}{\rho}\frac{\partial}{\partial \rho} +\frac{\partial^{2}}{\partial z^{2}} 
\end{equation*}

( $\overrightarrow{\rho}_{0}$ and $\overrightarrow{z}_{0}$ being unit vectors), i.e., they are similar to the ordinary  Laplacian and gradient operators for flat space expressed in cylindrical coordinates provided that there is no angular coordinate dependence.

It should be noted that the integrability condition of equations \ref{3.4}, \ref{3.5}  for determining the function  $\gamma(\rho, z)$  is the system \ref{3.2}, \ref{3.3} which does not contain    $\gamma(\rho, z)$ .  

 If we switch from the rotation potential $\omega$ to the new potential  $\Phi$  by the formulae 

\begin{equation}
    \frac{\partial \omega}{\partial \rho}=\frac{\rho}{f^2}\frac{\partial \Phi}{\partial z}, \quad \frac{\partial \omega}{\partial z}=-\frac{\rho}{f^2}\frac{\partial \Phi}{\partial \rho}, \label{3.6}
\end{equation}
we obtain the following field equations:

\begin{equation}
    f\Delta f=(\overrightarrow{\nabla}f)^2-(\overrightarrow{\nabla}\Phi)^2, \quad \overrightarrow{\nabla}(f^{-2}\overrightarrow{\nabla}\Phi)=0. \label{3.7}
\end{equation}

Introducing the Ernst complex potential 

\begin{equation}
    \varepsilon=f+i\Phi \label{3.8}
\end{equation}
gives us, from \ref{3.7},  the equation 

\begin{equation}
    (\varepsilon+\varepsilon^{*})\Delta \varepsilon=2(\overrightarrow{\nabla}\varepsilon)^2,\label{3.9}
\end{equation}
where $\varepsilon^{*}$  is the complex conjugate of  $\varepsilon$.

\section{The euclidon method}
A study of methods of generating exact solutions, based on internal symmetries of the Einstein equations, has been attracting the attention of gravitation physicists. 

The publications by Ehlers \cite{28} , Ozvath \cite{28} and Harrison \cite{29} pioneer these investigations. Subsequently, one can distinguish three approaches to this problem. 

The group-theoretic technique, which is very useful in construction of the new metrics containing an arbitrary large number of parameters, was introduced by Geroch \cite{30}, Kinnersley \cite{31} and Maison \cite{32}, and used in the papers \cite{33}-\cite{47}. The main achievements of this approach are related with the group of continuous transformations of the Ernst equation, with the Hoenselaers-Kinnersley-Xanthopoulos (HKX) transformations \cite{40},\cite{41}, with the aid of which one can obtain a set of asymptotically flat stationary vacuum metrics \cite{48}-\cite{70}. 

The second approach, introduced originally by Belinsky and Zakharov \cite{71}, 
 \cite{72}, is based on the application of the inverse scattering method to the Einstein's equations. At this point one can obtain new results in general relativity \cite{73}-\cite{92} . 

Finally, the last approach to generate new solutions from old was founded on the applications of the Backlund transformations. For the first time in general relativity the Backlund transformations were introduced by Harrison \cite{93} and Neugebauer \cite{94}, and their possible usage one can found from \cite{95}-\cite{103}.

Although these approaches are independent, they all are mathematically equivalent, that was shown by Cosgrove \cite{104}-\cite{106}. 

The major result of the applications of techniques discussed above is the construction of the non-linear superposition of N Kerr solutions aligned along their common rotational axis. Therewith the Tomimatsu-Sato metric has been recognized to be a special case of this solution \cite{107}-\cite{127}. 

Furthermore, the stationary vacuum problem can be reduced to the one singular equation \cite{128}. The application of the technique of the generalized Riemannian problem to general relativity yields a large class of new solutions \cite{129}.

In the paper \cite{130} the method of the variation of constants was proposed. With
the help of this technique one can brought to the nonlinear superposition of the Kerr space-time, with an arbitrary vacuum field \cite{131}, \cite{132}. 

In \cite{1} the nonlinear superposition of the stationary euclidon solution with an arbitrary axially symmetric stationary gravitational field on the basis of the method of variation of parameters was constructed. In the paper 
  \cite{2} stationary soliton solution of the Einstein equations was generalized to the case of a stationary seed metric. The formulae obtained in \cite{1}, \cite{2}  have a simple and compact form, permitting an effective nonlinear "addition" of the solutions. The euclidon method propounded in \cite{1}  was used in the article \cite{161}. 

In this paper the both methods \cite{1}, \cite{2}  are considered again and some new applications are presented. 

 One of the simplest solutions of equation  \ref{3.6}, \ref{3.7} is the one-stationary euclidon solution \cite{2}
 
\begin{equation*}
    f =\frac{f_{1:z_i}}{C_1}= \frac{(z - z_i) + r_i \tanh U_0}{C_{1}},
\end{equation*}

\begin{equation}
    \Phi =\frac{\Phi_{1:z_i}}{C_1}+C_2= \frac{r_i}{C_1 \cosh U_0} + C_2, \label{4.1}
\end{equation}

\begin{equation*}
    \omega =C_1\omega_{1:z_i}+C_3= C_1 \frac{r_i}{(z - z_i) + r_i \cdot \tanh U_0}  \cdot \frac{1}{\cosh U_0} + C_3,
\end{equation*}
where 
 \begin{equation}
    \quad r_{i} \equiv \left[\rho^2 + (z-z_{i})^2\right]^{1/2}; i=1,2,... \label{4.2}
\end{equation}

Here $z_i,U_{0}=U_{1:z_i}, C_1$, $C_2$, and $C_3$ are arbitrary constants. This is a solution that belongs to the Lewis's class of solutions \cite{4} .

It makes sense to call the solution \ref{4.1} a "statinary euclidon", since direct calculations show for corresponding metric \cite{2} all components of the Riemann-Christoffel curvature tensor turn to zero. It follows that the space-time is flat, which makes it possible to obtain transition formulae \cite{2} from the Minkowski interval. This fact provides slightly different perspective on the physical interpretation of well-known solutions, such as the Schwarzschild and Kerr solutions \cite{2}.

In order to make a nonlinear "composition" of the solution \ref{4.1} with the seed solution $(f^0(\rho, z)$,  $\Phi^0(\rho,z)$, $\omega^0(\rho,z))$, we use the method of variation of parameters. We shall consider $C_{1}$, $C_{2}$, $C_{3}$ and $U_{0}$ in the solution \ref{4.1} to be functions 

\begin{equation}
    C_1 \rightarrow f^0(\rho, z), \quad C_2 \rightarrow \omega^0(\rho, z), \quad C_3 \rightarrow \Phi^0(\rho, z), \quad U_0 \rightarrow U(\rho, z). \label{4.3}
\end{equation}

In this case we have

\begin{equation*}
    f = \frac{(z - z_i) + r_i \tanh U(\rho, z)}{f^0}, 
\end{equation*}

\begin{equation*}
    \Phi=\frac{r_i}{f^0\cdot \cosh U(\rho,z)}+\omega^0(\rho,z),
\end{equation*}

\begin{equation}
    \omega = f^0 \frac{r_i}{(z - z_i) + r_i\tanh U(\rho, z)}  \cdot \frac{1}{\cosh U(\rho, z)} + \Phi^0(\rho, z).\label{4.4}
    \end{equation}

A substitution of \ref{4.4} into equations \ref{3.6}, \ref{3.7} leads to the following set of first-order differential equations for the unknown function $U(\rho, z)=U_{z_i,f^0}(\rho, z):$

\begin{equation*}
    r_i \frac{\partial U}{\partial \rho} =  \frac{(z - z_i)}{f^0} \frac{\partial f^0}{\partial \rho} + \frac{\rho}{f^0}\frac{\partial f^0}{\partial z}+
    \end{equation*}
\begin{equation*}
    + \left[(z - z_i) \sinh U + r_i \cdot \cosh U \right] \frac{1}{f^0} \frac{\partial \Phi^0}{\partial \rho} +  \rho \frac{\sinh U}{f^0}\frac{\partial \Phi^{0}}{\partial z}, 
\end{equation*}

\begin{equation*}
    r_i \frac{\partial U}{\partial z} = -\frac{\rho}{f^0} \frac{\partial f^0}{\partial \rho}  +\frac{(z-z_i)}{f^0} \frac{\partial f^0}{\partial z}
    +
\end{equation*}

\begin{equation}
   +  \left[ (z - z_i) \sinh U +  r_i \cdot\cosh U \right] \frac{1}{f^0} \frac{\partial \Phi^0}{\partial z}-\rho\frac{\sinh U}{f^{0}}\frac{\partial \Phi^{0}}{\partial \rho}. \label{4.5}
\end{equation}

It is readily seen that the integrability condition for equations \ref{4.5} is satisfied.

Equations  \ref{4.5} are nonlinear. The problem of linearization of the system \ref{4.5} can be solved by the substitution $U(\rho, z) = \ln [a(\rho, z)/b(\rho, z)]$ \cite{2}.

 \section{ Two-stationary euclidon (soliton) solution (the Kerr-NUT solution)}
\textbf{a)}  Let us consider superposition  of the stationary  own-euclidon solution 
 ($f_{1:z_1},\ 
 \ \Phi_{1:z_1},\ \ \omega_{1:z_1}$)  whith the ($f^0=f_{1:z_2},\ 
 \ \Phi^0=\Phi_{1:z_2},\ \ \omega^0=\omega_{1:z_2}$) by using \ref{4.4}.

The function $U$ from \ref{4.5} in this case has the form \cite{2}

\begin{equation*}
   U(\rho, z)= U_{2:z_1,z_2} =U(\rho, z;z_1,z_2,\alpha,W)
   \end{equation*}
   where
   \begin{equation*}
   U(\rho, z;z_1,z_2,\alpha,W)=
   \end{equation*}
   \begin{equation*}
   =\ln[  (r_1+r_2+z_1-z_2) \sqrt{1 + \tanh W} -\alpha (z_1-z_2-r_2+r_1)\sqrt{1 - \tanh W}]-
\end{equation*}
 \begin{equation}
   -\ln[  \alpha(r_2-r_1+z_1-z_2) \sqrt{1 + \tanh W}+(z_1-z_2-r_2-r_1)\sqrt{1 - \tanh W}] \label{5.1}
\end{equation}
and $W=U_{1:z_2}=U_0,\  \ \alpha=\alpha_0$ and $\alpha_0$ is an integration constant.

For further generalizations it is convenient to put
\begin{equation}
    \tanh U_0 =\tanh U^a_0= \frac{1 - a_0^2}{1 + a_0^2}, \quad \alpha_0 =\varepsilon_2 \frac{b_0 - a_0}{1 + a_0 b_0},\quad \varepsilon_2=\pm1 \label{5.2}
\end{equation}

 In this case we get two-stationary euclidon solution (soliton) in the form of the Ernst complex potential \ref{3.8}

  \begin{equation}
    \varepsilon_{2:z_1,z_2} =f_{2:z_1,z_2}+i\Phi_{2:z_1,z_2}=   \frac{r_2 A_0 + r_1 B_0 + z_2-z_1}{r_2 A_0 + r_1 B_0 + z_1-z_2}. \label{5.3}
    \end{equation}
    $A_0$ and $B_0$ are determined by the relations 

 \begin{equation}
    A_0 = \frac{1 + ia_0}{1 - ia_0}, \quad B_0 = \frac{1 + ib_0}{1 - ib_0}, \label{5.4}
\end{equation}

If we put $z_1=-z_2=k_0$, then \ref{5.3} reduce to the Kerr-NUT solution \cite{2} in  the prolate ellipsoidal coordinates $(x,y)$: 

\begin{equation}
    \rho=k_{0}\sqrt{(x^2-1)(1-y^2)}, \quad z=k_{0}xy  \ (\ k_{0}= const) \label{5.5}
\end{equation}
and ($\varepsilon_2=1$)

\begin{equation*}
    \frac{1+a_{0}b_{0}}{\sqrt{(1+a^2_{0})(1+b^2_{0})}}=k_{0}(m^2_{0}+l^2_{0})^{-1/2}, \frac{b_{0}-a_{0}}{\sqrt{(1+a^2_{0})(1+b^2_{0})}}=a(m^2_{0}+l^2_{0})^{-1/2},
\end{equation*}

\begin{equation}
    \frac{1-a_{0}b_{0}}{\sqrt{(1+a^2_{0})(1+b^2_{0})}}=m_{0}(m^2_{0}+l^2_{0})^{-1/2}, \frac{b_{0}+a_{0}}{\sqrt{(1+a^2_{0})(1+b^2_{0})}}=-l_{0}(m^2_{0}+l^2_{0})^{-1/2}, \label{5.6}
\end{equation}
 we can obtain the Kerr-NUT   solution in Boyer-Lindquist coordinates $(r,\theta) \ \  (x=(r-m_{0})/k_{0}, \ \ y= cos\theta)$ 

\begin{equation*}
    ds^2=\frac{r^2+(a\cos \theta-l_{0})^2}{r^2-2m_{0}r+a^2-l^2_{0}}dr^2+[r^2+(a\cos\theta-l_{0})^2]\times
\end{equation*}

\begin{equation*}
    \times[d\theta^2+\frac{(r^2-2m_{0}r+a^2-l^2_{0})\sin^2 \theta d\varphi^2}{r^2-2m_{0}r+a^2\cos^2 \theta -l^2_{0}}]-[1-2\frac{m_{0}r+l_{0}(l_{0}-a\cos\theta)}{r^2+(a\cos \theta-l_{0})^2}]\times
\end{equation*}

\begin{equation}
    \times\{dt-(\frac{2a\sin^2 \theta[m_{0}r+l_{0}(l_{0}-a\cos \theta)]}{r^2-2m_{0}r+a^2\cos^2\theta-l^2_{0}}-2l_{0}\cos \theta)d\varphi\}^2. \label{5.7}
\end{equation}

If we set $l_{0}=0 \ \ (a_0 = -b_0)$, then (\ref{5.7}) yields the Kerr solution.

\textbf{b)} It turns out possible to generalize the two-stationary euclidon (soliton) solution \ref{5.3} with an arbitrary stationary vacuum field $\varepsilon^0=f^0+i\Phi^0$  in the Weyl canonical coordinates $(\rho, z)$. For this purpose it is better to use  a single relation for a complex function $\varepsilon=f+i\Phi$, i.e. \cite{2}
     
 \begin{equation}
    \varepsilon =\frac{(\varepsilon^0 -\varepsilon^{0*})}{2} +  \frac{(\varepsilon^0 + \varepsilon^{0*})}{2} \cdot \frac{r_2 A + r_1 B + z_2-z_1}{r_2 A + r_1 B + z_1-z_2}, \label{5.8}
    \end{equation}

If $f^0=1, \  \Phi^0=0$, then from \ref{5.8} we get the two-stationary euclidon solution \ref{5.3}.

$A$ and $B$ are determined by the relations 

 \begin{equation}
    A = \frac{1 + ia}{1 - ia}, \quad B = \frac{1 + ib}{1 - ib}, \label{5.9}
\end{equation}
and the functions $ a= \exp{(-U^a)} , \ \ b=-\exp{U^b}$  should be found from the first-order differential equations \ref{4.5} and  $U^a=U_{z_2,f^0}(\rho, z),\ \ U^b=U_{z_1,f^0}(\rho, z)$.

\section{N-center solution }

\textbf{a)}  Let us consider superposition  of the stationary  own-euclidon solution  whith the static Zipoy-like solution which is a superposition of 2n static euclidon (soliton) solutions to an arbitrary degree 

\begin{equation}
    f^0=f^n_{Z-l.}= \prod_{i=1}^{2n}[(z - Z_i) + R_i]^{(-1)^{i+1}\gamma_i},\quad \Phi^0=\omega^0=0. \label{6.1}
\end{equation}
where 
 \begin{equation*}
    \quad R_{i} \equiv \left[\rho^2 + (z-Z_{i})^2\right]^{1/2}.
\end{equation*}

In this case using \ref{4.4}, \ref{4.5}
\begin{equation*}
    f=f_{1:z_1;n} = \frac{(z - z_1) + r_1 \tanh U}{f^n_{Z-l.}},
\end{equation*}

\begin{equation*}
    \Phi=\Phi_{1:z_1;n} =\frac{r_1}{f^n_{Z-l.}\cdot \cosh U},
\end{equation*}

\begin{equation}
    \omega = \omega_{1:z_1;n}=\frac{f^n_{Z-l.}r_1}{(z - z_1) + r_1\tanh U}  \cdot \frac{1}{\cosh U}  \label{6.2}
    \end{equation}
    where
    
\begin{equation}
   U=U_{1:z_1;n}(\rho, z) = \sum^{2n}_{i=1}(-1)^{i+1}\gamma_i\ln\frac{R_i+r_1+z_1-Z_i}{R_i-r_1+z_1-Z_i}+C, \label{6.3}
\end{equation}
$C$ is an integrating constant.

If $n=0 \ \ (f^n_{Z-l.}=1) $ we get from \ref{6.2}, \ref{6.3} the stationary  own-euclidon solution \ref{4.1}

\textbf{b)} Let us consider superposition  of the stationary  own-euclidon solution \ref{4.4} with the static own-euclidon solution $(1-\delta_0) $-th degree

\begin{equation}
    f^0= [(z - z_1) + r_1]^{1-\delta_0},\quad \Phi^0=\omega^0=0. \label{6.4}
\end{equation}

In this case we will get in a sense  "stationary euclidon solution $\delta_0 $-th degree"  of the form
\begin{equation*}
    f=f^{\delta_0}_{1:z_1} = \frac{(z - z_1) + r_1 \tanh U}{[(z - z_1) + r_1]^{1-\delta_0}},
\end{equation*}
\begin{equation*}
    \Phi=\Phi^{\delta_0}_{1:z_1}=\frac{r_1}{[(z - z_1) + r_1]^{1-\delta_0}\cdot \cosh U},
\end{equation*}

\begin{equation}
    \omega = \omega^{\delta_0}_{1:z_1}=\frac{[(z - z_1) + r_1]^{1-\delta_0}\sqrt{\rho^2 + (z - z_1)^2}}{(z - z_1) + r_1\tanh U}  \cdot \frac{1}{\cosh U}  \label{6.5} 
    \end{equation}
 where
\begin{equation}
   U=U_{1:z_1}^{\delta_0}(\rho, z) = (1-\delta_0)\ln\frac{r^2_1}{(z - z_1) + r_1}+U_0, \label{6.6}
\end{equation}

  In the static limit $  U_0 \rightarrow \infty $ we'll get the static own-euclidon solution $\delta_0 $-th degree
  
\begin{equation}
    f^{\delta_0}_{1:z_1} \rightarrow  [(z - z_1) + r_1]^{\delta_0},\ \
    \Phi^{\delta_0}_{1:z_1} \rightarrow 0,\ \ \omega^{\delta_0}_{1:z_1}\rightarrow 0. \label{6.7}
\end{equation}

If $\delta_0=1$ we get from \ref{6.5}, \ref{6.6} the stationary  own-euclidon solution \ref{4.1}

\textbf{c)}  Let us consider superposition  of the stationary  own-euclidon solution \ref{4.1}
 ($f_{1:z_1},\ 
 \ \Phi_{1:z_1},\ \ \omega_{1:z_1}$) with \ref{6.2} ($f_{1:z_2;n},\ 
 \ \Phi_{1:z_2;n},\ \ \omega_{1:z_2;n}$) using \ref{5.8}, \ref{5.9}

\begin{equation*}
    f=f_{2:z_1,z_2;n} = \frac{(z - z_1) + r_1 \tanh U}{f_{1:z_2;n}}=
\end{equation*}
\begin{equation*}
    =f^n_{Z-l.}  \frac{(z - z_1) + r_1 \tanh U}{(z - z_2) + r_2 \tanh U_{1:z_2;n}},
\end{equation*}
\begin{equation*}
    \Phi=\Phi_{2:z_1,z_2;n} =\frac{r_1}{f_{1:z_2;n}\cdot \cosh U}+\omega_{1:z_2;n},
\end{equation*}

\begin{equation}
\omega=\omega_{2:z_1,z_2;n}=\frac{f_{1:z_2;n}\cdot r_1}{(z - z_1) + r_1\tanh U}  \cdot \frac{1}{\cosh U} + \Phi_{1:z_2;n} \label{6.8}
\end{equation}
where using \ref{5.1}

\begin{equation*}
   U(\rho, z)=U_{2:z_1,z_2;n}=U(\rho, z;z_1,z_2,\alpha,W),\ \ W=U_{1:z_2;n}
 \end{equation*}
 
 \begin{equation}
    \alpha=\alpha_{2:z_1,z_2;n}=\varepsilon_2 \frac{\exp{U_{1:z_1;n}}+\exp{(-U_{1:z_2;n}})}{\exp{U_{1:z_1;n}}\exp{(-U_{1:z_2;n}})-1}, \  \ \varepsilon_2=\pm1.  \label{6.9}
\end{equation}

If $n=0 \ \ (f^n_{Z-l.}=1) $ we get the stationary two-euclidon solution \ref{5.3}.

\textbf{d)}  Let us consider superposition  of the stationary  own-euclidon solution 
 ($f_{1:z_1},\ 
 \ \Phi_{1:z_1},\ \ \omega_{1:z_1}$)  whith the ($f_{2:z_2,z_3;n},\ 
 \ \Phi_{2:z_2,z_3;n},\ \ \omega_{2:z_2,z_3;n}$) by the same method as in \cite{2}
    
\begin{equation*}
    f=f_{3:z_1,z_2,z_3;n} = \frac{(z - z_1) + r_1 \tanh U}{f_{2:z_2,z_3;n}},
\end{equation*}

\begin{equation*}
    \Phi=\Phi_{3:z_1,z_2,z_3;n} =\frac{r_1}{f_{2:z_2,z_3;n}\cdot \cosh U}+\omega_{2:z_2,z_3;n},
\end{equation*}
 
\begin{equation}
\omega_{3:z_1,z_2,z_3;n}=\frac{f_{2:z_2,z_3;n}\cdot r_1}{(z - z_1) + r_1\tanh U}  \cdot \frac{1}{\cosh U}  +\Phi_{2:z_2,z_3;n} \label{6.10}
\end{equation}
where using \ref{5.1}

    \begin{equation*}
   U(\rho, z)=U_{3:z_1,z_2,z_3;n}=U(\rho, z;z_1,z_2,\alpha,W),\ \ W=U_{2:z_2,z_3;n}
 \end{equation*}
 
 \begin{equation}
 \alpha=\alpha_{3:z_1,z_2,z_3;n}=\varepsilon_3 \frac{\exp{U_{2:z_1,z_3;n}}+\exp{(-U_{2:z_2,z_3;n}})}{\exp{U_{2:z_1,z_3;n}}\exp{(-U_{2:z_2,z_3;n}})-1}, \  \ \varepsilon_3=\pm1. \label{6.11}
\end{equation}

\textbf{e)}  Let us consider by induction superposition  of the stationary  own-euclidon solution 
 ($f_{1:z_1},\ 
 \ \Phi_{1:z_1},\ \ \omega_{1:z_1}$)  whith the ($f_{k-1:z_2,z_3,...,z_k;n},\ 
 \ \Phi_{k-1:z_2,z_3,...,z_k;n},\ \ \omega_{k-1:z_2,z_3,...,z_k;n}$)
 
\begin{equation*}
    f=f_{k:z_1,z_2,...,z_k;n} = \frac{(z - z_1) + r_1 \tanh U}{f_{k-1:z_2,z_3,...,z_k;n}},
\end{equation*}

\begin{equation*}
    \Phi=\Phi_{k:z_1,z_2,...,z_k;n} =\frac{r_1}{f_{k-1:z_2,z_3,...,z_k;n}\cdot \cosh U}+\omega_{k-1:z_2,z_3,...,z_k;n},
\end{equation*}
 
\begin{equation}
\omega=\omega_{k:z_1,z_2,...,z_k;n}=\frac{f_{k-1:z_2,z_3,...,z_k;n}\cdot r_1}{(z - z_1) + r_1\tanh U}  \cdot \frac{1}{\cosh U} +\Phi_{k-1:z_2,z_3,...,z_k;n}, \label{6.12}
\end{equation}
where using \ref{5.1}

 \begin{equation*}
   U(\rho, z)=U_{k:z_1,z_2,...,z_k;n}=U(\rho, z;z_1,z_2,\alpha,W),\ \ W=U_{k-1:z_2,z_3,...,z_k;n},
 \end{equation*}
 
 \begin{equation}
\alpha=\alpha_{k:z_1,z_2,...,z_k;n}=\varepsilon_k \frac{\exp{U_{k-1:z_1,z_3,...,z_k;n}}+\exp{(-U_{k-1:z_2,z_3,...,z_k;n}})}{\exp{U_{k-1:z_1,z_3,...,z_k;n}}\exp{(-U_{k-1:z_2,z_3,...,z_k;n}})-1}, \  \ \varepsilon_k=\pm1. \label{6.13}
\end{equation}

  It is obvious that under the condition $k=2N,\ \ n=N,\ \  z_i=Z_i \ \ (r_i=R_i),\ \ \gamma_{2l-1}=\gamma_{2l} \ \ (i=2l-1)$ in \ref{6.1} the solution \ref{6.12}, \ref{6.13} is a N-center solution. In this case from \ref{6.3}

  \begin{equation*}
   U=U_{1:z_1;n}(\rho, z) = \sum^{2n}_{i=1}(-1)^{i+1}\gamma_i\ln\frac{r_i+r_1+z_1-z_i}{r_i-r_1+z_1-z_i}+C=
   \end{equation*}
   
   \begin{equation}
   =\sum^{2n}_{i=1}(-1)^{i+1}\gamma_i\ln\frac{(r_i+r_1+z_1-z_i)^2}{2(z_1-z_i)(z-z_i+r_i)}+C, \label{6.14}
\end{equation}
$C$ is an integrating constant.

 If $  i=1\ \ (n=\frac{1}{2}),\ \ \gamma_1= 1-\delta_{0}$ from \ref{6.14} we get \ref{6.6}.
 
  \section{Two-center solution }
  
If $ k=2N=4,\ \ n=2,\ \  z_i=Z_i, \ \  \gamma_1=\gamma_2=\delta_{01}-1,\ \ \gamma_3=\gamma_4=\delta_{02}-1,$ up to a permutation of indices from \ref{6.12} we get two-center solution

\begin{equation*}
    f=f_{4:z_3,z_4,z_1,z_2;2} = \frac{(z - z_3) + r_3 \tanh U_{4:z_3,z_4,z_1,z_2;2}}{f_{3:z_4,z_1,z_2;2}}=
\end{equation*}
\begin{equation*}
     =[\frac{(z - z_3) + r_3 }{(z - z_4) + r_4}]^{\delta_{02}-1} \frac{(z - z_3) + r_3 \tanh U_{4:z_3,z_4,z_1,z_2;2}}{(z - z_4) + r_4 \tanh U_{3:z_4,z_1,z_2;2}} \times
     \end{equation*}
     \begin{equation*}
    \times [\frac{(z - z_1) + r_1 }{(z - z_2) + r_2}]^{\delta_{01}-1} \frac{(z - z_1) + r_1 \tanh U_{2:z_1,z_2;2}}{(z - z_2) + r_2 \tanh U_{1:z_2;2}},
\end{equation*}

\begin{equation*}
    \Phi=\Phi_{4:z_3,z_4,z_1,z_2;2.} =\frac{r_3}{f_{3:z_4,z_1,z_2;2}\cdot \cosh U_{4:z_3,z_4,z_1,z_2;2}}+\omega_{3:z_4,z_1,z_2;2},
\end{equation*}
 
\begin{equation}
\omega=\omega_{4:z_3,z_4,z_1,z_2;2}=\frac{f_{3:z_4,z_1,z_2;2}\cdot r_3}{(z - z_3) + r_3\tanh U_{4:z_3,z_4,z_1,z_2;2}}  \cdot \frac{1}{\cosh U_{4:z_3,z_4,z_1,z_2;2}} + \label{7.1} 
    \end{equation}
\begin{equation*}
    +\Phi_{3:z_4,z_1,z_2;2}. 
\end{equation*}

\textbf{a)} In the static limit
\begin{equation*}
\tanh U_{4:z_3,z_4,z_1,z_2;2}=\tanh U_{3:z_4,z_1,z_2;2}=\tanh U_{2:z_1,z_2;2}=\tanh U_{1:z_2;2}=1
\end{equation*}
we get   two-Zypoy solution

\begin{equation}
    f= [\frac{(z - z_1) + r_1}{(z - z_2) + r_2}]^{\delta_{01}}[\frac{(z - z_3) + r_3}{(z - z_4) + r_4}]^{\delta_{02}} ,\ \ \omega=\Phi=0 \label{7.2}
\end{equation}

If we put $\delta_{01}=1, \ \ \delta_{02}=0,\ \ z_1=-z_2=m$, then \ref{7.2} reduces to the Schwarzschild solution \cite{2}.

\textbf{b)} If $\delta_{01}=\delta_{02}=1$ from \ref{7.1} we get two-Kerr-NUT solution

\begin{equation*}
    f=f_{4:z_3,z_4,z_1,z_2} = \frac{(z - z_3) + r_3 \tanh U_{4:z_3,z_4,z_1,z_2}}{f_{3:z_4,z_1,z_2}}=
\end{equation*}
\begin{equation*}
     = \frac{(z - z_3) + r_3 \tanh U_{4:z_3,z_4,z_1,z_2}}{(z - z_4) + r_4 \tanh U_{3:z_4,z_1,z_2}} \times
      \frac{(z - z_1) + r_1 \tanh U_{2:z_1,z_2}}{(z - z_2) + r_2 \tanh U_{1:z_2}},
\end{equation*}

\begin{equation*}
    \Phi=\Phi_{4:z_3,z_4,z_1,z_2.} =\frac{r_3}{f_{3:z_4,z_1,z_2}\cdot \cosh U_{4:z_3,z_4,z_1,z_2}}+\omega_{3:z_4,z_1,z_2},
\end{equation*}

\begin{equation}
\omega=\omega_{4:z_3,z_4,z_1,z_2}=\frac{f_{3:z_4,z_1,z_2}\cdot r_3}{(z - z_3) + r_3\tanh U_{4:z_3,z_4,z_1,z_2}}  \cdot \frac{1}{\cosh U_{4:z_3,z_4,z_1,z_2}} + \label{7.3}
    \end{equation}
\begin{equation*}
    +\Phi_{3:z_4,z_1,z_2}. 
\end{equation*}

 If $z_3 \to z_4$, then \ref{7.3} reduces to the Kerr-NUT solution \ref{5.3}

 \begin{equation}
   f_{4:z_3,z_4,z_1,z_2} \to f_{2:z_1,z_2} \ \ \Phi_{4:z_3,z_4,z_1,z_2} \to \Phi_{2:z_1,z_2}, \  \ \omega_{4:z_3,z_4,z_1,z_2} \to \omega_{2:z_1,z_2} \label{7.4}
\end{equation}

\textbf{c)} If $z_3=z_4,\  \delta_{02}=1,\ \ \delta_{01}=\delta \  (n=1) $ we get "rotating Zipoy mass"

\begin{equation*}
    f_{2:z_1,z_2;1} = \frac{(z - z_1) + r_1 \tanh U_{2:z_1,z_2;1}}{f_{1:z_2;1}}
    = [\frac{(z - z_1) + r_1 }{(z - z_2) + r_2}]^{\delta-1}  \cdot \frac{(z - z_1) + r_1 \tanh U_{2:z_1,z_2;1}}{(z - z_2) + r_2 \tanh U_{1:z_2;1}},
\end{equation*}

\begin{equation*}
    \Phi_{2:z_1,z_2;1} =\frac{r_1}{f_{1:z_2;1}\cdot \cosh U_{2:z_1,z_2;1}}+\omega_{1:z_2;1},
\end{equation*}

\begin{equation}
    \omega_{2:z_1,z_2;1}=\frac{f_{1:z_2;1}\cdot r_1}{(z - z_1) + r_1\tanh U_{2:z_1,z_2;1}}  \cdot \frac{1}{\cosh U_{2:z_1,z_2;1}}  +\Phi_{1:z_2;1}. \label{7.5}
\end{equation}

In case  $\delta=1 \ \ (n=0) $, \ref{7.5} tends to \ref{5.3}. 

In the static limit
\begin{equation*}
\tanh U_{2:z_1,z_2;1}=\tanh U_{1:z_2;1}=1
\end{equation*}
from \ref{7.5} we get Zypoy solution \cite{2}.

With the coordinate transformation \ref{5.5} and \ref{5.6} $ (a_0 = -b_0)$ the asympyotic behavior ($r \to \infty$) of \ref{7.5} in Boyer-Lindquist coordinates $(r,\theta) \ \  (x=(r-m_{0})/k_{0}, \ \ y= cos\theta)$ takes the form

\begin{equation}
f_{2:k_0,-k_0;1} \to 1-\frac{2\delta m}{r}, \ \ \omega_{2:k_0,-k_0;1} \to \frac{2\delta J \sin^2{\theta}}{r}, \label{7.6}
\end{equation}
where $J=ma$ is the Kerr angular momentum.

Tomimatsu-Sato solution \cite{136} for deformation parametr $\delta=2$ has the same asymptotics \ref{7.6}.

\section{ Euclidon algebra} \label{8}

From \ref{6.8} follows the law of composition

\begin{equation}
    f_{2E:i,k;N}(U_{2E:i,k;N})=f_{1E:i} \times f^{-1}_{1E:k;N}= 
\begin{pmatrix}
f^{-1}_{1:k;N}(U_{1:k;N})f_{1,i}(U_{2:i,k;N})\\
f^{-1}_{1:k;N}(U_{1:k;N})\Phi_{1:i}(U_{2:i,k;N})+\omega_{1:k;N}(U_{1:k;N}) \\
f_{1:k;N}(U_{1:k;N})\omega_{1:k;N}(U_{2:i,k;N})+\Phi_{1:k;N}(U_{1:k;N}) 
\end{pmatrix} \label{8.1}
 \end{equation}
where

\begin{equation}
   f_{1E:k;N}= f_{1E:z_k;N}(U_{1:z_k;N}) = 
\begin{pmatrix}
f_{1:z_k;N}(U_{1:z_k;N})\\
\Phi_{1:z_k;N}(U_{1:z_k;N}) \\
\omega_{1:z_k;N}(U_{1:z_k;N}) 
\end{pmatrix}. \label{8.2}
 \end{equation}

 In case  $N=0,\ \ (1E:k;N)=(1E:k)$, \ref{8.1}, \ref{8.2} tend to the law of composition of two stationary euclidons \cite{3}.

It's easy to see that \cite{2}

\begin{equation*}
    f_{1E:i}(U_{1:i}) \times f^{-1}_{1E:i}(U_{1:i})= e
\end{equation*}
where
\begin{equation}
   f_{1E:i}= f_{1E:z_i}(U_{1:z_i}) = 
\begin{pmatrix}
f_{1:z_i}(U_{1:z_i})\\
\Phi_{1:z_i}(U_{1:z_i}) \\
\omega_{1:z_i}(U_{1:z_i}) 
\end{pmatrix} \label{8.3}
 \end{equation}
and single element

\begin{equation}
    e = 
\begin{pmatrix}
1\\
0 \\
0 
\end{pmatrix}. \label{8.4}
 \end{equation}

For the inverse element we have
\begin{equation}
    e\times f^{-1}_{1E:i}(U_{1:i})= f^{-1}_{1E:i}(U_{1:i})=
\begin{pmatrix}
f^{-1}_{1:i}(U_{1:i})\\
\omega_{1:i}(U_{1:i}) \\
\Phi_{1:i}(U_{1:i}) 
\end{pmatrix}. \\ \label{8.5}
 \end{equation}

 It's easy to see that
\begin{equation}
     (f^{-1}_{1E:i}(U_{1:i}))^{-1}= f_{1E:i}(U_{1:i}) \label{8.6}
\end{equation}
 and
 \begin{equation}
    (f_{1E:i}(U_{1:i}) \times f^{-1}_{1E:k}(U_{1:k}))^{-1}= f^{-1}_{1E:i}(U_{1:i})\times f_{1E:k}(U_{1:k})= f_{1E:k}(U_{1:k})\times f^{-1}_{1E:i}(U_{1:i}). \label{8.7}
\end{equation}

Expressions \ref{6.10} can be obtained in the following two ways (associativity condition)
\begin{equation*}
    f_{3E:i,k,l;N}=(f_{1E:i} \times f^{-1}_{1E:k})\times f_{1E:l;N}=  f_{1E:i} \times (f_{1E:k} \times f^{-1}_{1E:l;N})^{-1}=
    \end{equation*}
\begin{equation}
    =
\begin{pmatrix}
f^{-1}_{1:k}(U_{2:k,l;N})f_{1:l;N}(U_{1:l;N})f_{1:i}(U_{3:i,k,l;N})\\
f^{-1}_{1:k}(U_{2:k,l;N})f_{1:l;N}(U_{1:l;N})\Phi_{1:i}(U_{3:i,k,l;N})+f_{1:l;N}(U_{1:l;N})\omega_{1:k}(U_{2:k,l;N})+\Phi_{1:l;N}(U_{1:l;N}) \\
f_{1:k}(U_{2:k,l;N})f^{-1}_{1:l;N}(U_{1:l;N})\omega_{1:i}(U_{3:i,k,l;N})+f^{-1}_{1:l;N}(U_{1:l;N})\Phi_{1:k}(U_{2:k,l;N})+\omega_{1:l;N}(U_{1:l;N}) 
\end{pmatrix}. \label{8.8}
\end{equation}

In \ref{8.8} it is taken into account that  $U_{3:i,k,l;N}=U_{3:i,(k,l;N)}=U_{3:(i,k),l;N}$ that are determined by recurrent formula \ref{6.13}.

\section{  Conclusion}
Thus, using the simple nonlinear superposition method  we got the stationary solution of Einstein’s vacuum
Equations  describing "N rotating axially symmetric masses", which in
the absence of rotation describes various N Zipoy masses \cite{137} on the axis of symmetry, and in the absence of distortion,
N Kerr-NUT masses.

The solution \ref{6.12} is an exact solution, but most likely, by its construction, it approximately describes N rotating Zipoy masses. This is evidenced, for example, by singularities on the event horizon of such solutions \cite{131}, \cite{138}. However, the solution \ref{6.12} is multicenter and takes distortion into account to some extent, while the solutions \cite{124}-\cite{126} are one-center generalizations of the Kerr solution in the presence of distortion  $\delta $.

Moreover, if instead of  the solution \ref{6.1} we take the next solution

\begin{equation*}
    f^0= \prod_{k=1}^{n}[\frac{(z - z_{2k-1}) + r_{2k-1}}{(z - z_{2k}) + r_{2k}}]^{-1} \cdot f^{static}_k ,\quad \Phi^0=\omega^0=0,
\end{equation*}
then the multicenter solution obtained in a similar way using Euclidon algebra \ref{8}. In the static limit will give a set of arbitrary known static solutions $f^{static}_k $ on the axis of symmetry. The presence of such sources "on a thread" is confirmed experimentally \cite{139}.
 
Using Bonnor's theorem \cite{140} for \ref{6.12} under the condition $k=2N,\ \ n=N,\ \  z_i=Z_i \ \ (r_i=R_i),\ \ \gamma_{2l-1}=\gamma_{2l}=-\frac{1}{2} \ \ (i=2l-1)$, it is easy to construct a solution describing N massive Gutsunaev-Manko magnetic dipoles \cite{131}  and reducing to various Schwarzschild masses \cite{2}  on the axis of symmetry in a static limit. 

Based on single center 2N-euclidon (soliton) solution \ref{6.12}  $(n=0,\ \
z_{2l-1}=-z_{2l}=m\ \ (i=2l-1) )$, in \cite{141} proposed a novel class of KERR-SEN solutions respecting the SO(2) symmetry group, systematically constructed via the Laurent series expansion technique. Based on these considerations in \cite{141} received the multiplet $(f,\Phi, \phi , \chi) $, 
where $ \phi  $ and $ \chi $ are dilaton and axion.

\section{Acknowledgments}

We would like to thank Professor H. Quevedo, Professor A. De Benedictis and Professor F. Frutos for helpful discussions.

\end{document}